# Accelerated Real-Life (ARL) Testing and Characterization of Automotive LiDAR Sensors to facilitate the Development and Validation of Enhanced Sensor Models

**Marcel Kettelgerdes[1,3], Tjorven Hillmann[3], Thomas Hirmer[3], Hüseyin Erdogan[2], Bernhard Wunderle[4] and Gordon Elger[1,3]**

(1) Technical University of Applied Sciences Ingolstadt (THI), Germany
(2) Conti Temic microelectronic GmbH, Germany
(3) Fraunhofer Institute for Transportation and Infrastructure Systems (IVI), Germany
(4) Chemnitz University of Technology (TUC), Germany

***Abstract*** *– In the realm of automated driving simulation and sensor modeling, the need for highly accurate sensor models is paramount for ensuring the reliability and safety of advanced driving assistance systems (ADAS). Hence, numerous works focus on the development of high-fidelity models of ADAS sensors, such as camera, Radar as well as modern LiDAR systems to simulate the sensor behavior in different driving scenarios, even under varying environmental conditions, considering for example adverse weather effects. However, aging effects of sensors, leading to suboptimal system performance, are mostly overlooked by current simulation techniques. This paper introduces a cutting-edge Hardware-in-the-Loop (HiL) test bench designed for the automated, accelerated aging and characterization of Automotive LiDAR sensors. The primary objective of this research is to address the aging effects of LiDAR sensors over the product life cycle, specifically focusing on aspects such as laser beam profile deterioration, output power reduction and intrinsic parameter drift, which are mostly neglected in current sensor models. By that, this proceeding research is intended to path the way, not only towards identifying and modeling respective degradation effects, but also to suggest quantitative model validation metrics.*

## Introduction

In the course of the ongoing mobility transformation towards electrified and automated driving, especially advanced driving assistance systems (ADAS) and automated driving (AD) functions are subject to a steep increase in complexity. Given the consequently enormous physical test effort required to ensure the reliable operation of those safety-critical systems under varying driving scenarios and environmental conditions, virtual validation becomes an indispensable tool within ADAS development and testing [Don22, Gal20].

Since, the reliable environmental perception serves as a basis for respective AD functions, such as automated lane keeping (ALKS) or cruise control (ACC), numerous works focus on the development of high-fidelity sensor models, including for example non-optical sensors such as Radar, as well as optical camera and LiDAR systems. Especially LiDAR sensors which are, due to their high spatial resolution and accuracy, widely regarded as a key enabler for highly automated driving [Roy19], are strongly investigated due to their sensitivity against adverse weather conditions, more specifically precipitation, fog, or direct sun irradiance. The respective measurement campaigns were generally either conducted in lab environments, in which fog and rain conditions are emulated [Hai,23, Mon21, Ras11] or in an outdoor setting, in which sensors are tested under real weather conditions via automated long-term measurements [Ket23, Lin22].

While numerous recent works already dealt with modelling and validating such effects [Hai,23, Ket23, Lin22], aging-related sensor performance degradation is mostly overlooked in current model development approaches. One major reason for this might lie in the lack of sensor-specific aging data, which again might occur from the fact that degradation data is generally regarded as sensitive from the manufacturer side and that the LiDAR sensors itself are still expensive, such that researchers are not willed to afford multiple sensors in order to age them in an accelerated manner. Another reason might be that there are yet no defined aging procedures for LiDAR sensors among the research community. However, reliable sensor performance has not only to be ensured under varying environmental conditions, but also over the whole product life cycle and therefore several years of operation. This is further underlined by the well-known aging effects of major LiDAR components, like the laser [Zha22], the receiver diodes [Kam22], or the scanning mechanism [Yoo22], which do finally not only motivate intelligent condition monitoring of the sensor during operation [Goe20, Ket21, Str19, Str20], but also to model its macroscopic system behaviour over lifetime. For both purposes, it is paramount to understand which degradation effects are encountered on system level and how they impact the performance of the sensor.

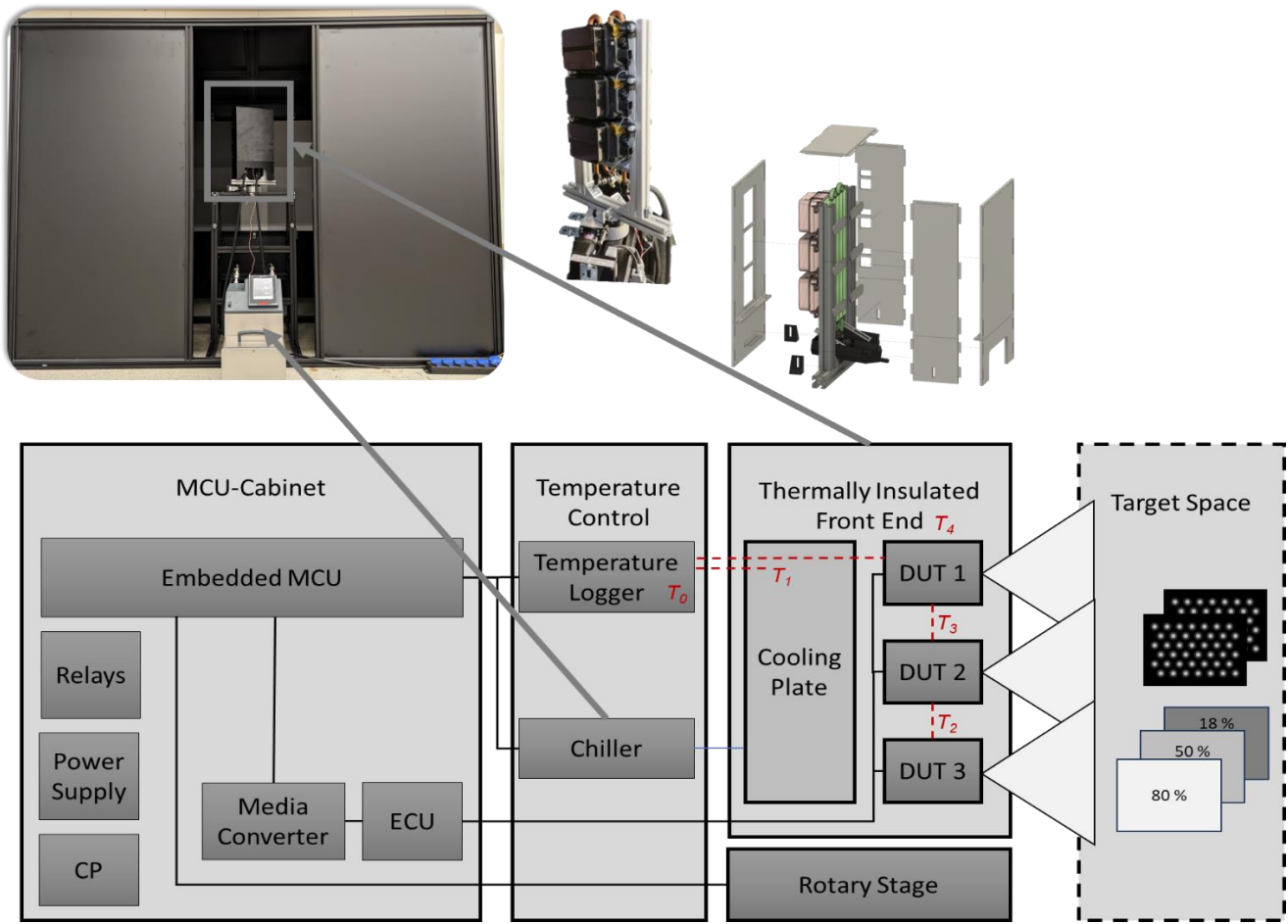

**Figure 1. Illustration of the LiDAR HiL hardware architecture. While the bottom side shows the functional block diagram of the overall system, the upper side contains a CAD explosion and photographic illustration of the respective components.**

Based on that, this paper introduces a cutting-edge Hardware-in-the-Loop (HiL) test bench designed for the parallel accelerated aging and characterization of three Automotive LiDAR sensors from Continental. The primary objective of this research is to address the aging effects of LiDAR sensors, specifically focusing on aspects such as laser beam profile deterioration and output power reduction, which are mostly neglected in current sensor models. The HiL test bench is equipped to perform accelerated aging experiments by subjecting the LiDAR sensors to operational temperature fluctuations within the range of -10 °C to 85 °C. This temperature control is achieved through the implementation of a chiller system and a vertically aligned cooling plate onto which the sensors are securely mounted. Notably, the cooling plate is affixed to a high-precision rotary stage, enabling controlled rotation of the sensors over Lambertian reflection targets and intrinsic recalibration targets. To eliminate stray light effects and ensure precise data acquisition, the entire setup is enclosed within a measuring cabinet. Continuous recording of sensor data throughout the temperature cycling process allows for the tracking and modelling of key performance parameters over its operational lifetime. These include laser beam profile, output power, the number of dead pixels, as well as the set of intrinsic calibration parameters. In summary, this study presents an innovative HiL test bench tailored for the accelerated aging and comprehensive characterization of LiDAR sensors. The research serves as a crucial step towards enhancing operating lifetime dependent sensor models for virtual validation, ultimately contributing to the advancement of safe and reliable automated driving systems in an increasingly complex automotive landscape.

The work is structured as follows. In the next section, the hardware (HW) and software (SW) design of the test bench, as well as the devise under test (DUT) including its data acquisition process, are described in detail. Moreover, the test sequence including the thermal load profile is introduced. In the third section the in-situ data processing is presented and investigated measurement effects are identified based on first sample data. Finally, the fourth section offers a conclusion of the contribution.

# ARL-HiL Testbench

For the accelerated real-life tests, a HiL test bench including hardware and ROS-based software was developed based on an in-situ sensor test methodology, after which the sensors are continuously operated while frequently gathering LiDAR detection data as well as internal data, such as temperatures, voltages, and currents of key components. While the detection data is recorded to study the aging impact on the sensor performance, internal data is gathered for a subsequent correlation analysis in order to investigate the potential for in-field condition monitoring.

## Hardware and Software Architecture

### Hardware architecture

The overall hardware setup is shown in Fig.1. The whole installation consists of a main computing unit (MCU) cabinet, a temperature control unit and a thermally insulated front end, which is placed in a target-equipped, stray light-enclosed measurement cabinet. In addition to the respective embedded PC, the MCU-cabinet contains an electronic control unit (ECU) to communicate with the sensor DUT, as well as media converters to translate the automotive BroadR-Reach signal from the ECU to a 100BASE-T ethernet signal which is forwarded to the MCU. Additionally, the main control cabinet contains the overall power supply, circuit protection (CP) as well as relays to software-wise toggle the power supply for major components like the DUT.

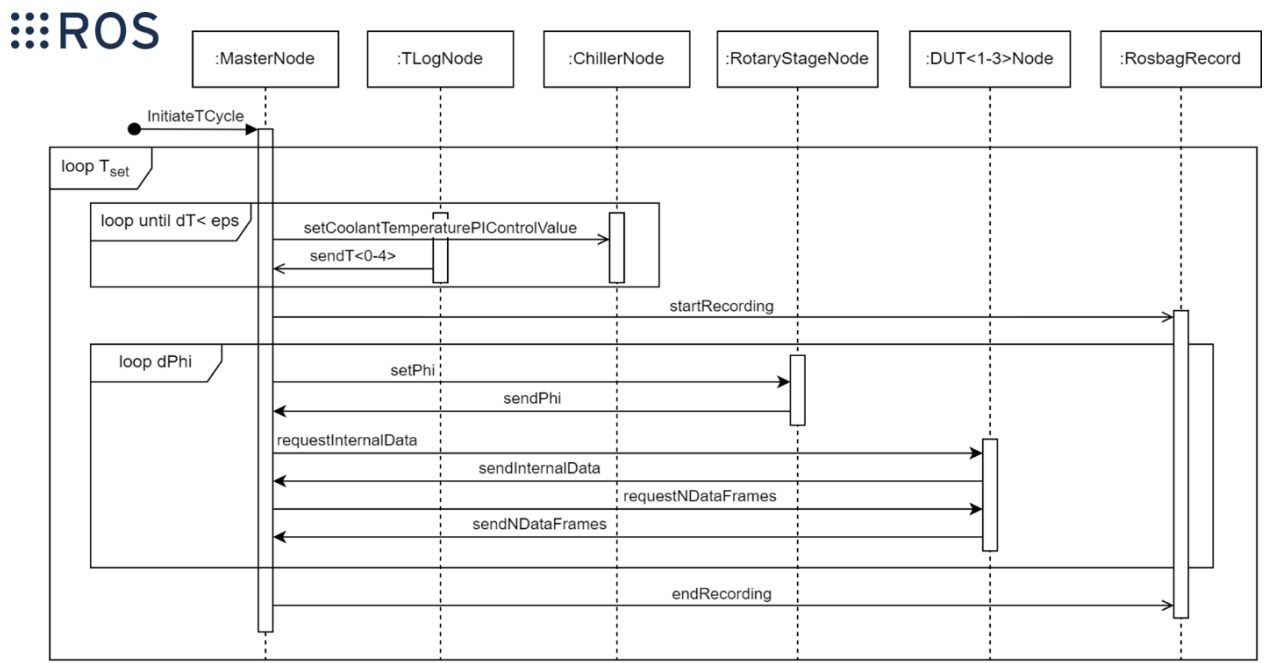

**Figure 2. ROS-based software architecture of the LiDAR HiL test bench as UML sequence diagram. Every key component is implemented as a ROS node which is communicating with a respective master node.**

The temperature control unit consists of a chiller from Huber, which is cooling (resp. heating) a silicon oil-based coolant to a given set temperature while pumping it over insulated hoses through an aluminium cooling plate. The internal set temperature of the coolant $T_{oil}$ is calculated based on a cascaded PI controller, which is implemented on the MCU, controlling the housing temperatures of the DUT ($T_2$-$T_4$) to a set temperature $T_{set}$. All externally measured temperatures ($T_0$-$T_4$) are measured by thermocouples, containing next to the DUT housing temperatures, the cooling plate temperature between inlet and outlet ($T_1$) and the ambient temperature $T_0$. The temperature logger is connected to the MCU via a serial connection, while the chiller is connected over 100BASE-T ethernet. While receiving coolant set temperatures, it sends back internal data, like coolant level, system warnings and actual coolant temperature.

The core of the measurement setup consists of the insulated front end, on which the three DUT are vertically aligned on the cooling plate and mounted using thermal interface material (TIM) with a heat conductivity of $\lambda = 6$ W/mK for improved heat transfer between DUT and cooling plate. The whole front end is enclosed in an ABS housing and insulated with an internal 6 mm thick elastomer-based insulation with a heat conductivity of $\lambda \leq 0,035$ W/mK (see Fig. 1, top). Furthermore, the housing contains a condensation drain, which enables drainage of the considerable amount of melting ice and condensation on the cooling plate and sensor housings at phase transition. The front end including enclosure is mounted on a rotary stage from Gunda-Automation, consisting of a high-precision stepper motor and the respective motor control. This enables a sub-degree accurate horizontal rotation, which again allows for a pixelwise rotation of the sensor frustums over a statically placed target. The rotary stage is connected over a serial connection with the MCU, receiving an absolute set angle and sending back the actual positioning angle. The target setup consists of two intrinsic calibration targets and three centrally aligned Lambertian reflectance stripes and will be described in detail in section 3.

## Software Architecture and Test Sequence

The software architecture was implemented based on the robot operating system (ROS), running on an Ubuntu 20 industrial embedded PC. Respectively, the drivers of all major components which are connected to the MCU, are implemented ROS nodes, which are asynchronously publishing and receiving node-specific messages to the ROS network. The main test procedure is illustrated in Fig.2 in form of a unified modeling language (UML) sequence diagram.

**Table 1. Investigated Sensor data.**

| Detection Data | Description |
|---|---|
| $I(u,v)$ | Pixelwise measured intensity of first laser return pulse |
| $D(u,v)$ | Pixelwise measured distance of first laser return pulse |
| **Operating Data** | **Description** |
| $T_{FPGA}$ | FPGA die temperature |
| $T_{ToF}$ | ToF die temperature |
| $T_{Laser}$ | Laser diode temperature |
| $V_{TECN}, V_{TECP}$ | TEC voltage on n- and p-side |
| $I_{TEC}$ | TEC current |
| $V_{Laser}$ | Voltage over laser diode |
| $I_{Laser}$ | Laser pulse current |
| $t_{pump}$ | Laser pump time |
| $V_{xVx}$ | All main power supply voltage rails (e.g., 1V8, 3V3 …) |

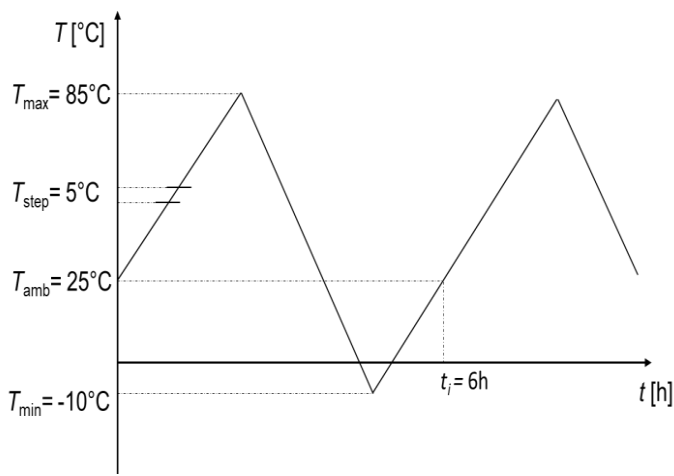

Figure 3. Thermal load profile of the DUT. The sensors are thermally cycled between $T_{min}$ = -10°C and $T_{max}$ = 85°C during operation, while in every $T_{step}$ = 5°C temperature step sensor data is recorded. One full cycle is conducted within $t_i$ = 6h.

Thus, the temperature cycling, and data recording are basically scheduled by a master node. For a given temperature load cycle (see Fig. 3), it loops in $T_{step}$ = 5 °C steps through the given load cycle (in this case from $T_{set}$ = -10 °C-85 °C) and sets it as set temperature for the PI temperature controller. As soon as all DUT housing temperatures $T_2$-$T_4$ have settled to $T_{set}$ within a thermocouple measurement tolerance ε, the master initiates data recording. After that, it rotates the front end clockwise over the target space by looping through the DUTs horizontal field of view (*HFoV*) by φ≈[-60°,+60°] and increasing the rotation angle of the rotary stage in steps of Δφ ≈ 0.9° (corresponding to the spatial resolution - and hence, a single pixel in the image space - of the sensors). Within every rotation step it requests all available internal data from the sensor (key component temperatures, voltages, and currents) as well as 10 raw intensity and distance frames. As soon as the set angle reaches the end position, the front end is rotated back to the initial position and recording is stopped and saved as a serialized rosbag with the respective temperature step, time stamp and cycle count.

## Device Under Test (DUT)

Coming to the sensor hardware, an automotive short-range Flash LiDAR system from Continental is investigated (see Fig. 1, top and Fig. 5, top right). It is a non-scanning direct time-of-flight sensor, which provides distance and intensity image frames D and I, as well as an equivalent point cloud PC, which can again be derived by applying the inverse intrinsic transformation ($K^{-1}$, $\delta^{-1}$) to D and I [Ket21, Ket23]. The distance and intensity data frames can be interpreted as raw detection data on which subsequent object detection algorithms extract the necessary environmental information needed for a specific AD function, like ALKS. Therefore, it allows to investigate the impact of aging on the sensor's detection performance at an early stage, in a way that specific measurement effects (see Sec. 3) can be traced back to respective sensor components.

The D/I-frames are generated by pixelwise evaluation of the backscattered laser pulse in the time domain [Roy19]. Thus, after emitting a laser pulse, a parallel ToF and laser intensity measurement is triggered in each pixel of the ToF imager (in this case with a resolution of W x H = 128 x 32 pix). Based on the resulting intensity-timeseries, the first two intensity peaks $i = \{1,2\}$ are detected as the first two significant pulse returns from surrounding objects. For each pixel (u,v) and return pulse $i$, the peak's maximum value is extracted as intensity value $I_i(u,v)$ and the respective ToF $t_i$ is used to calculate the corresponding object distance $D_i(u,v)$ based on the known speed of light c:

$$D_i(u,v) = c \cdot t_i / 2 \qquad (1)$$

In case of the DUT, the fast FPGA-based return pulse computation finally achieves a sensor frame rate of up to 25 Hz. It operates on a laser wavelength of 1064 nm with a field of view of HFoV x VFoV = 120° x 27.5°, covering a measurement range from $D$ = 0.5 m to 25 m. It is worth mentioning that, within this work, only the first LiDAR return pulse is considered in each pixel ($I(u,v) = I_1(u,v)$, $D(u,v) = D_1(u,v)$), since multipath effects within the measurement chamber can be neglected.

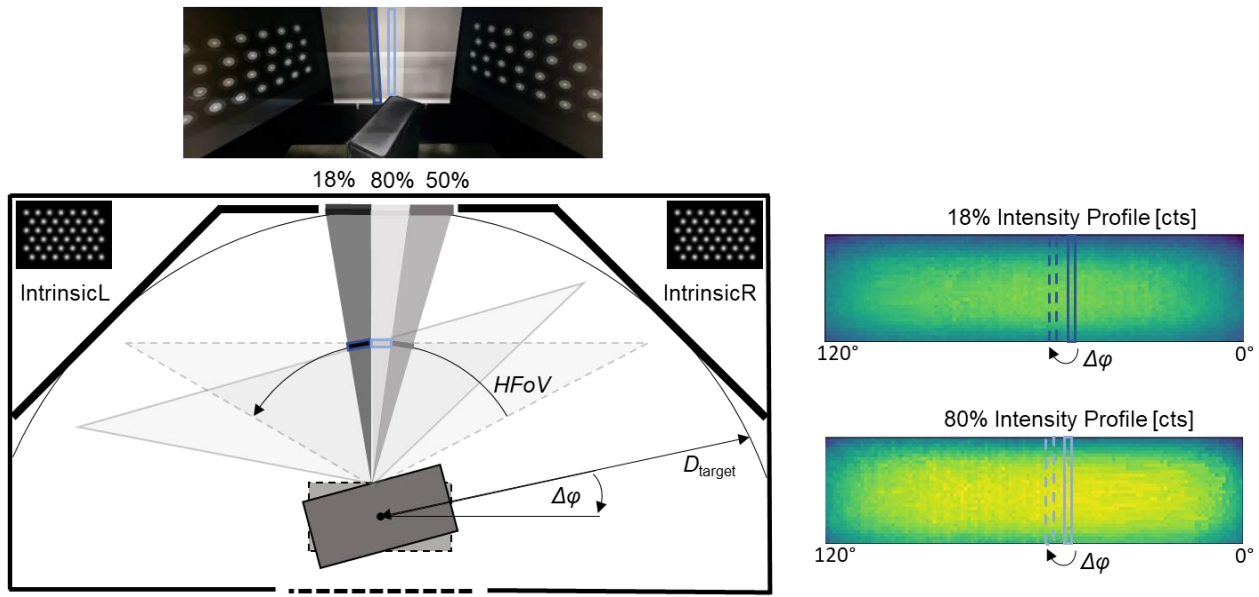

**Figure 4. Target setup and data acquisition process. The DUT frustums are rotated over two on a semicircle aligned intrinsic calibration targets on the left and right side, as well as over three Lambertian reflectance stripes of 18 %, 50 % and 80 % reflectance to calculate the sensors received intensity profile for varying reflectances, by which again the laser beam profile in form of the relative spatial optical output power distribution can be derived.**

## Sensor Data Acquisition

In addition to the above-described raw detection data, internal operating data of the sensor (see Tab.1, bottom) is tracked and rather investigated with respect to a potential in-field condition monitoring implementation, than for sensor modeling. This data comprises the operating temperatures of all major components, such as the DUT's ToF imager ($T_{ToF}$), FPGA ($T_{FPGA}$) and laser unit ($T_{Laser}$). Moreover, the operating voltages and currents of the sensor's major loads, in form of the laser diode ($I_{Laser}$, $V_{Laser}$) and its thermo-electric cooler (TEC), as well as all main power supply voltage rails ($V_{xVx}$), are recorded. In sum, all these thermal and electrical operating parameters complement the detection data in a way that correlation analyses between the cheap, in-field accessible operation data and the lifetime dependent sensor performance are facilitated. Following this, it could for instance be investigated whether an imminent failure of the laser can be detected by an anomaly detection based on the laser current $I_{Laser}$, pump time $t_{pump}$, temperature $T_{Laser}$ and voltage $V_{Laser}$. Furthermore, a direct correlation with the laser output power could enable continuous power monitoring or even a remaining useful lifetime prediction. However, in order to measure the sensor's performance degradation quantitatively, detection data-based performance metrics are required, which will be discussed in the next section.

# Detection Data Processing and Evaluation

As indicated in the previous section, during data acquisition, the DUT are rotated over statically placed targets, which are aligned on a semicircle around the rotated front end, ensuring that the central area of the three targets has the same distance $D_{target}$ = 1.1 m (see Fig.4). The left (IntrinsicL) and right (IntrinsicR) target are installed for continuous intrinsic re-calibration of the DUT in order to track and assess the intrinsic calibration parameters $K$ as a compact representation of the sensor front end's opto-mechanical state [Ket21]. For that, both targets are fully contained in the rotation angle ranges of φ≈[-60°,-30°] and φ≈[+30°,+60°], respectively. This results in a full frustum occupancy of the intrinsic calibration targets, which is required to facilitate a reliable and re-producible calibration. Since, in addition to that, for each target angle 10 intensity frames are recorded, a temporal mean intensity image can be computed for each target angle to reduce pixel noise impact onto the feature point detection and hence, improve calibration accuracy [Ket21].

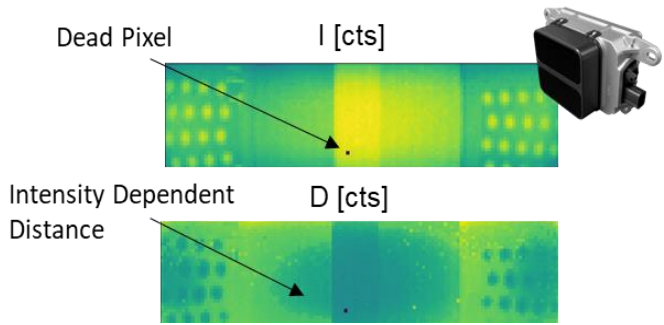

**Figure 5. Sample data of the DUT during rotation, consisting of a 12-bit intensity frame I and the corresponding distance frame D. Exemplary measurement effects are marked.**

**Table 2. Investigated Measurement effects.**

| Key Metric | Description |
|---|---|
| $P_o(φ,θ)$ | Pixel- or ray-direction-wise rel. deviation of Intensity |
| $ΔK$ | Rel. deviation of intrinsic parameters |
| $N_{dead}$ | Dead pixel count |
| $σ_I(u,v)$ | Pixelwise temporal intensity STD |
| $σ_D(u,v)$ | Pixelwise temporal distance STD |
| $ΔD(u,v)$ | Pixelwise rel. distance accuracy |

The central target contains three vertical Lambertian reflectance stripes of 18 %, 50 % and 80 % reflectance. By rotating the DUT frustums over a fixed region of interest (ROI) on all three reflectance targets (see Fig. 4, in blue), the respective backscattered laser intensity can be measured for varying incident angles and a fixed target distance. This allows to generate an angle- (or pixel-) dependent intensity profile across the sensor frustum for all three target reflectances, exemplarily shown for the 18 % and 80 % target ROI in Fig. 4 on the right. Under the assumption, that the sensor's optical efficiency of the receiver front end is spatially invariant, this offers the possibility to measure the relative optical output power distribution $P_o(φ,θ)$ of the laser (often referred to as beam profile) over lifetime. It is worth mentioning that the targets would ideally be spherically scanned, since in a pure horizontal rotation, the ray distance will be slightly higher in the upper and lower frustum section, compared to the central section, but since the half-angle VFoV /2 is solely 13.75°, the corresponding maximum ray distance variation for $D_{target}$ = 1.1 m is neglected with $ΔD_{target}$ ≤0.27m. This is further backed by the fact that this and subsequent research focuses rather on the aging-dependent relative deviation of the beam profile and not an accurate absolute measurement of the output power.

In addition to the transmitter's optical output power distribution and the receiver's intrinsic calibration parameters, further, more accessible key effects are investigated as listed in Tab.2. These comprise the number of dead pixels $N_{dead}$, which is requested as part of the internal sensor data but can also be identified in single sensor frames (see Fig. 5). Moreover, the distance measurement accuracy $\Delta D(u,v)$ with respect to the ground truth target distance and backscattered intensity is evaluated. As one can see in Fig. 5, the distance measurement accuracy is strongly intensity dependent, which is resulting from the impact of the return signal strength, respectively the pulse height, onto the peak detection performance. Lastly, the pixel noise in form of the pixelwise temporal standard deviation (STD) $\sigma(u,v)$ in the respective distance and intensity frame is investigated with respect to temperature and aging induced drift.

# Conclusion

In conclusion, this paper introduces a Hardware-in-the-Loop test bench designed for accelerated aging and characterization of LiDAR sensors. The proceeding research is dedicated to address the aging effects of LiDAR sensors, focusing on optical performance on receiver and transmitter side. By subjecting the sensors to extreme temperature fluctuations, this study offers first insights into potentially critical parameters and their impact on sensor performance. Therefore, it constitutes a crucial step towards enhancing lifetime-dependent sensor models for virtual validation, contributing to the advancement of safe and reliable automated driving systems in an increasingly complex automotive landscape. Subsequent works will present the results of the aging tests including model approaches for key degradation effects, as well as a correlation analysis with internal operating currents, voltages, and temperatures in order to investigate the potential for in-field condition monitoring.

# Acknowledgement

This work is done in the context of the projects: "KI-LiDAR – Miniaturisierte LIDAR-Sensoren mit KI-Zustandsüberwachung für das autonome Fahren" and "Gaia-X 4 PLC-AAD – Product Life Cycle – Across Automated Driving". The financial support of the German federal ministry of education and research (BMBF) and the German federal ministry of Economic Affairs and Climate Action (BMWK), which have made the current work possible, is greatly acknowledged.